\documentstyle[epsfig]{mn}
\begin{document}
\def\ltsima{$\; \buildrel < \over \sim \;$}
\def\simlt{\lower.5ex\hbox{\ltsima}}
\def\gtsima{$\; \buildrel > \over \sim \;$}
\def\simgt{\lower.5ex\hbox{\gtsima}}
\def\approxgt{\mathrel{\hbox{\rlap{\lower.55ex \hbox {$\sim$}}
        \kern-.3em \raise.4ex \hbox{$>$}}}}
\def\approxlt{\mathrel{\hbox{\rlap{\lower.55ex \hbox {$\sim$}}
        \kern-.3em \raise.4ex \hbox{$<$}}}}

\title[]{IGR J17354$-$3255 as a candidate intermediate SFXT possibly associated with the transient MeV AGL J1734$-$3310}

\author[Sguera et al.]
{V. Sguera$^{1,2}$, S.P. Drave$^{3}$, A.J.Bird$^{3}$, A. Bazzano$^2$, R. Landi$^1$,  P. Ubertini$^2$
\\
$^1$ INAF, Istituto di Astrofisica Spaziale e Fisica Cosmica, Via Gobetti 101, I-40129 Bologna, Italy \\
$^2$ INAF, Istituto di Astrofisica Spaziale e Fisica Cosmica, Via Fosso del Cavaliere 100, I-00133 Rome, Italy\\
$^3$ School of Physics and Astronomy, University of Southampton, University Road, Southampton, SO17 1BJ, UK \\
}

\date{Accepted 2011 June 21. Received in original form 2011 March 2}

\maketitle
{}

\begin{abstract}
We present spectral and temporal results from INTEGRAL long-term monitoring of the unidentified X-ray source IGR J17354$-$3255.
We show that it is a weak persistent hard X-ray source spending  a major fraction of the time in an
out-of-outburst state with average 18--60 keV X-ray flux of $\sim$ 1.1 mCrab, occasionally interspersed 
with fast X-ray flares (duration from a few hours  to a few days) with a dynamic range as high as $\sim$ 200. 
From archival \emph{Swift}/XRT observations, we also show that the dynamic range from non-detection 
to highest level of measured X-ray activity is $>$300. Our IBIS timing analysis strongly confirms the $\sim$ 8.4 days 
orbital period previously detected with \emph{Swift}/BAT, in addition we show that the
shape of the orbital profile  is rather smooth and appears to
be dominated by low level X-ray emission rather than by bright outbursts, the measured  
degree of outburst recurrence is $\sim$ 25\%. The spectral and temporal characteristics of 
IGR J17354$-$3255 are highly indicative of a Supergiant High Mass X-ray Binary nature (SGXB). 
However, our inferred dynamic ranges both at soft and hard X-rays are significantly greater than those 
of classical SGXB systems, but instead 
are typical of intermediate Supergiant Fast X-ray Transient (SFXTs). 
Finally, we note for the first time that the observed fast flaring X-ray behaviour of 
IGR J17354$-$3255 is very similar to that detected with AGILE
from  the spatially associated  MeV source  AGL J1734$-$3310, suggesting
a possible physical link between the two objects. 

\end{abstract}

\begin{keywords}
X-rays:binaries -- X-rays: individual (IGR J17354-3255)
\end{keywords}

\vspace{1.0cm}

\section{INTRODUCTION}
IGR J17354$-$3255 is an unidentified  hard X-ray transient discovered with INTEGRAL  on April 2006 
(Kuulkers et al. 2006, 2007). Its average flux at that time was $\sim$ 18 mCrab (20--60 keV) and no information  
is available on the duration of its X-ray activity because it was in the IBIS/ISGRI field of view (FOV) 
for only a few hours. The source is located in the direction of the Galactic Center,
a region extensively monitored by INTEGRAL during the last $\sim$ 8 years. As a result, IGR J17354$-$3255  is reported in the latest 4th IBIS catalog 
(Bird et al. 2010) with a 18--60 (20--40) keV average flux of $\sim$ 1.7 mCrab or 2.2$\times$10$^{-11}$ 
erg cm$^{-2}$ s$^{-1}$ (1.4 mCrab)  and on-source exposure of $\sim$ 7.6 Ms.  
It is also listed in the 54 months \emph{Swift}/BAT hard X-ray catalogue (Cusumano et al. 2010) with 
a very similar average flux of 2.1$\times$10$^{-11}$ erg cm$^{-2}$ s$^{-1}$ (15--150 keV).   
The \emph{Swift} broad band spectrum (0.2--100 keV) is typical  of accreting neutron stars in High Mass 
X-ray Binaries (HMXBs), i.e. an absorbed cutoff power law  (D'Ai et al. 2011). In addition, timing analysis of the \emph{Swift}/BAT light  
curve showed modulation at $\sim$8.4 days which could be interpreted as the likely  orbital period of 
the binary system (D'Ai et al. 2011).

In the soft X-ray band (0.2--10 keV), to date the region of the sky including 
IGR J17354$-$3255 was observed twice by \emph{Swift}/XRT,  on 2008 and 2009 (Vercellone et al. 2009, D'Ai et al. 2011). 
Only during the observation on April  2009 was a single X-ray counterpart detected inside 
the $\sim$1.4 arcminutes IBIS/ISGRI error circle, suggesting a transient nature. 
This X-ray source  was also detected by Chandra on February 2009 (Tomsick et al. 2009) at a flux level 
of $\sim$ 1.3$\times$10$^{-11}$ erg cm$^{-2}$ s$^{-1}$, i.e. similar to that also measured by \emph{Swift}/XRT.
The Chandra 0.2--10 keV spectrum  was rather hard ($\Gamma$ $\sim$ 0.5) and intrinsically absorbed (N$_H$ $\sim$ 7$\times$10$^{22}$ cm$^{-2}$),  
supporting a HMXB nature for IGR J17354$-$3255. It is worth noting that only one 
bright 2MASS counterpart is located inside the refined Chandra error circle (0.2 arcseconds radius) while there are no catalogued USNO optical objects (Tomsick et al. 2009). The high inferred optical extinction and the relatively high X-ray column density suggest that the source 
is probably  distant and located at least at the distance of the Galactic Center region ($\sim$8.5 kpc). 

Finally, the position of IGR J17354$-$3255 is spatially correlated with that of AGL J1734$-$3310,  an unidentified
transient MeV source discovered by AGILE on April 2009 during a flare lasting only one day (Bulgarelli et al. 2009). 
Although the hypothesis of a  physical link between the two objects  is intriguing, 
the large positional uncertainty of AGL J1734$-$3310 (error radius of 0$^\circ$.65) makes this association
rather circumstantial and  further in-depth studies are needed.

Here we report for the first time on detailed spectral and timing analysis of INTEGRAL data 
of IGR J17354$-$3255. In particular, the INTEGRAL long-term monitoring allowed us to discover 15 new hard X-ray flares 
and  properly study the out-of-outburst X-ray emission. Our findings are crucial to understand the nature of the source and the 
origin of its X-ray emission. Finally we used all the collected informations to investigate the possibility of a 
physical association between IGR J17354$-$3255 and AGL J1734$-$3310 and we discuss any possible major implication stemming 
from this association.

\section{DATA ANALYSIS}
For the INTEGRAL study, we used all the public data collected with IBIS/ISGRI (Ubertini et al. 2003, Lebrun et al. 2003) 
and JEM-X (Lund et al. 2003)  from the end of February 2003 to the end of October  2008. 
In particular, the IBIS (JEM-X) data set  consists of $\sim$ 7050 ($\sim$ 602) pointings or Science Windows 
(ScWs, $\sim$ 2000 seconds duration) where IGR J17354$-$3255 was within 12$^\circ$ (5$^\circ$)
from the centre of the instruments FOV.  A 12$^\circ$ limit was applied because the off-axis response of 
IBIS/ISGRI is not well modelled at large off-axis angles and in combination with the telescope dithering 
(or the movement of the source within the FOV) it may introduce a systematic error in the measurement 
of the source fluxes.  IBIS/ISGRI images for each pointing were generated in the energy band
18--60 keV using  the ISDC offline scientific  analysis software version 7.0. 
Count rates at the position of the source were extracted from all individual 
images to produce its long  term light curve on the ScW timescale.  

In addition, we also used  data collected with the BAT and XRT instruments on board the \emph{Swift}  
satellite (Gehrels et al. 2004). From  the \emph{Swift}/XRT archive,  IGR J17354$-$3255
was observed on March 2008 and April 2009. Here we present a spectral and temporal analysis, the XRT data 
reduction was  performed  according to the processes described in Landi et al. (2010). 
From  the \emph{Swift}/BAT archive, 
we downloaded the 15--50 keV light curve  of IGR J17354$-$3255  on daily timescale. 
After filtering for poor quality data flags, short exposures and low coded aperture fractions, 
the \emph{Swift}/BAT light curve covers from April 2009 through  May 2011 with an effective exposure time
of  $\sim$ 6.4 Ms. 

All spectral analysis was performed  using \emph{Xspec}  version 11.3; 
uncertainties are given at the 90\% confidence level for one single parameter of interest.

\begin{table}
\caption {Summary of all IBIS detections of hard X-ray flares from IGR J17354$-$3255. The table lists the date 
of their peak emission, approximate duration  and significance detection of the entire flaring activity,  X-ray flux  at the peak, power law photon index with   
$\chi^{2}_{\nu}$ and  d.o.f., and finally reference to the discovery paper of each flare, i.e. (1) this paper and  (2) Kuulkers et al 2006; $\ddagger$ = upper limit on the duration,  $\star$  = lower limit on the duration.}
\label{tab:main_outbursts} 
\begin{tabular}{ccccccc}
\hline
N. & peak-date    &  duration    & sig    &  peak-flux       & $\Gamma$                   & ref \\
    &  (MJD)     &   (hours)     &  $\sigma$        & (18--60 keV)     &($\chi^{2}_{\nu}$, d.o.f.)  &      \\
   &            &                &                  &   (mCrab)        &                            &      \\
\hline                                                                            
1  & 52741.5  & $\sim$65$\ddagger$  & 9.0     & 25$\pm$5        & 2.0$^{+0.7}_{-0.7}$ (1.2,15)                   & 1   \\
2  & 53051.9 &$\sim$0.5             & 5.5   & 108$\pm$20        & 2.2$^{+1.0}_{-2.0}$ (1.2,15)        & 1   \\
3  & 53114.9  &$\sim$10             & 5.5   & 35$\pm$12       & 2.3$^{+2.0}_{-2.0}$ (1.2,15)             & 1   \\
4  & 53452.4  &$\sim$0.5            & 4.2   & 25$\pm$6        &                           & 1   \\
5  & 53602.9  &$\sim$0.5            & 4.4   & 35$\pm$8        &                           & 1   \\
6 & 53794.6  &$\sim$0.5             & 5.0     & 25$\pm$5        & 2.6$^{+2.0}_{-2.5}$ (1.3,15)    & 1   \\
7 & 53813.8  &$\sim$60$\ddagger$    & 5.6   & 27$\pm$6        & 2.2$^{+1.6}_{-1.6}$  (0.9,15)       & 1   \\
8 & 53829.8  &$\sim$36              & 6.6   & 30$\pm$4        & 2.6$^{+0.8}_{-0.8}$  (0.8,15)      & 1   \\
9 & 53846.2  &$\sim$5$\star$        & 8.0   & 28$\pm$6        & 2.0$^{+0.7}_{-0.7}$   (0.8,15)      & 2   \\
10 & 53975.3  &$\sim$3              & 4.6   & 32$\pm$7        &                           & 1   \\
11 & 53999.7  &$\sim$1.5            & 5.2   & 30$\pm$7        & 2.1$^{+1.2}_{-1.2}$ (1.1,15)      & 1   \\
12 & 54012.6  &$\sim$5              & 4.2   & 40$\pm$9        &                          & 1   \\
13 & 54340.2  &$\sim$5              & 6.5   & 25$\pm$7        & 2.5$^{+2.0}_{-2.0}$  (1.01,15)      & 1   \\
14 & 54345.5  &$\sim$63             & 7.0  & 21$\pm$6        & 3.2$^{+1.0}_{-1.0}$ (1.2,15)       & 1   \\
15 & 54539.1  &$\sim$25             & 6.2   & 21$\pm$5        & 2.1$^{+1.5}_{-2.0}$ (1.2,15)          & 1   \\
16 & 54547.8  &$\sim$3$\star$       & 5.7   & 30$\pm$6        & 2.9$^{+1.0}_{-1.0}$ (0.9,15)        & 1   \\
\hline
\hline  
\end{tabular}
\end{table}

\section{TEMPORAL ANALYSIS}
\subsection{IBIS/ISGRI}
\subsubsection{Light curve}
We performed a detailed investigation of the 18--60 keV IBIS/ISGRI long-term light curve on
ScW timescale of IGR J17354$-$3255 (total on source time $\sim$ 10 Ms) 
to fully characterize for the first time its temporal behaviour. 
We found that most of the time the source is not significantly detected at ScW level (2$\sigma$ upper limit of $\sim$10 mCrab)
and it is well below the instrumental 
sensitivity of IBIS/ISGRI.
This is very likely the reason why the source  was not reported in the 2nd IBIS catalog (Bird et al. 2006) 
despite the region of the sky was observed for  a total  exposure time of $\sim$1.6 Ms.
On the contrary,  the source is  listed in the subsequent 3th and 4th IBIS catalogs (Bird et al. 2007, 2010) 
with a similar 18--60 (20--40)  keV average flux of $\sim$ 1.7 mCrab (1.4 mCrab); this  is  very likely due to the significantly 
longer time intervals analyzed  in the 3th ($\sim$ 4.5 Ms) and 4th IBIS catalog ($\sim$ 7.6 Ms)
with respect to the 2nd one.

In addition, we also investigated the IBIS light curve in order to unveil a possible flaring behaviour
of IGR J17354$-$3255. To this aim, we only considered those flares having a peak-flux greater than $\sim$ 20 mCrab or 
2.6$\times$10$^{-10}$ erg cm$^{-2}$ s$^{-1}$ (18--60 keV); 
we adopted such  conservative peak-flux threshold for flare recognition for two main reasons: i) it corresponds to a source significance 
equal to or greater than $\sim$4$\sigma$ in the single ScW containing 
the peak of the flare, ii) to pick up flares bright  enough to extract a meaningful ISGRI spectrum.
By applying this criterion, a total of 16  X-ray flares have been detected 
over a total exposure of $\sim$ 115 days though not in sequence, as listed in Table 1 together 
with the date of the peak emission, approximate duration  and significance detection of the entire outburst activity,  
flux at  the peak.  We note that all but one  are newly discovered flares and are reported for the first time, 
their typical  duration is only a few hours (0.5--5 h), however the source 
occasionally displays X-ray activity over a period of a few days (1--3 d). The typical peak-flux 
is in a narrow range $\sim$ 20--30 mCrab  but also brighter flares occur ($\sim$ 35--45 mCrab).
The strongest as well as shortest  outburst (N.2 in Table 1) was detected only during one
ScW lasting half an hour. Such fast X-ray transient behaviour is 
evident from Fig. 1 and 2 which  show  the IBIS/ISGRI light curve on ScW timescale and the corresponding sequence of 
consecutive ScWs significance images, respectively. In particular, the inset in Fig. 1 represents a zoomed view
with a bin time of 200 seconds, i.e. significantly smaller than that at ScW timescale. The outburst is characterized by two fast flares lasting a few minutes, the strongest one reached a peak-flux of 
108$\pm$20 mCrab  or (1.41$\pm$0.26)$\times$10$^{-9}$ erg cm$^{-2}$ s$^{-1}$.

\begin{figure}
\epsfig{file = 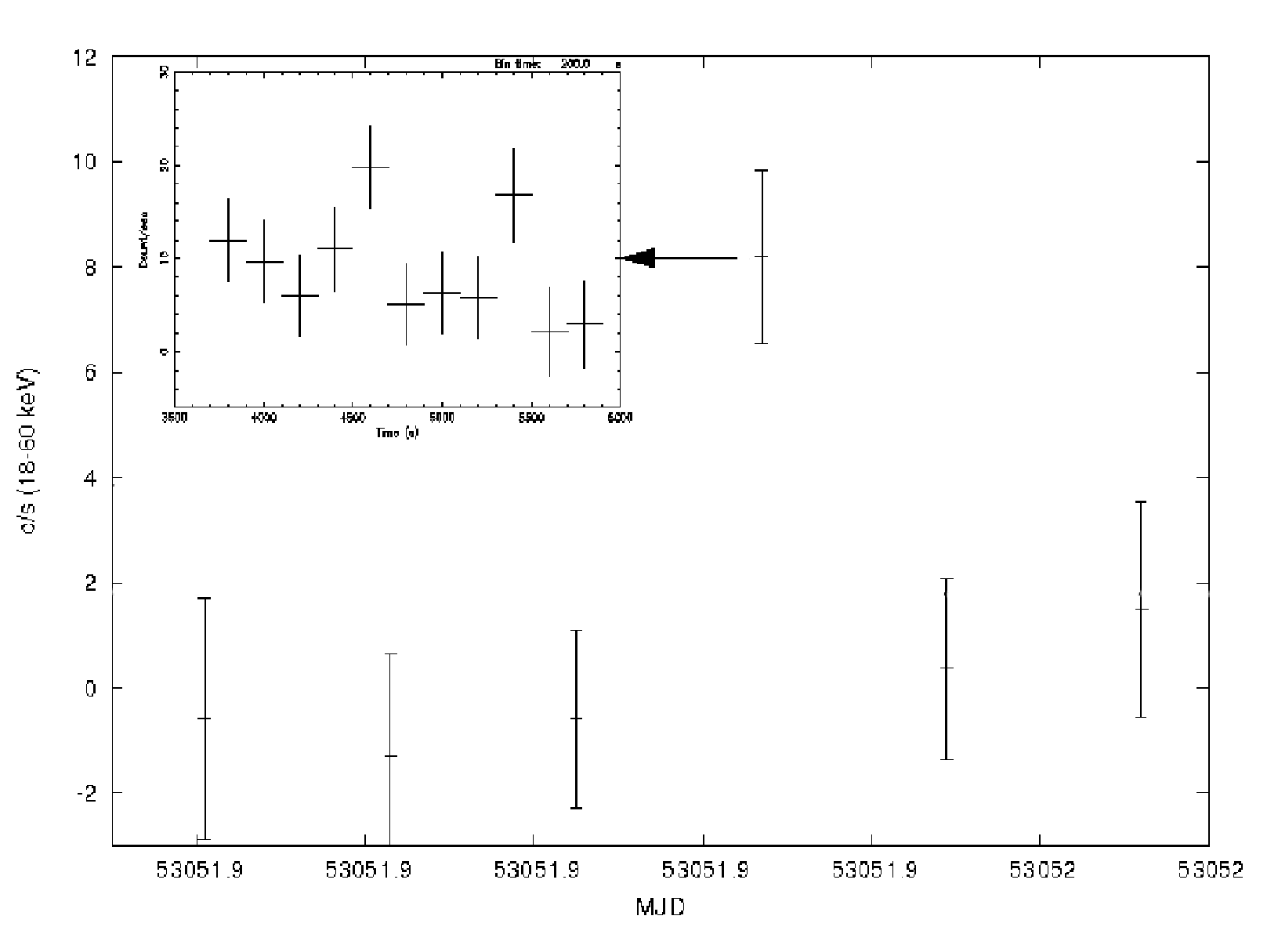, height=7cm,width=8.5cm}
\caption{IBIS/ISGRI light curve (18--60 keV) of flare N.2 in Table 1. Each data point represents the average flux 
during one ScW ($\sim $2000 seconds bin time). The inset represents a zoomed view of the flare with a bin time of 200 seconds.} 
\epsfig{file = 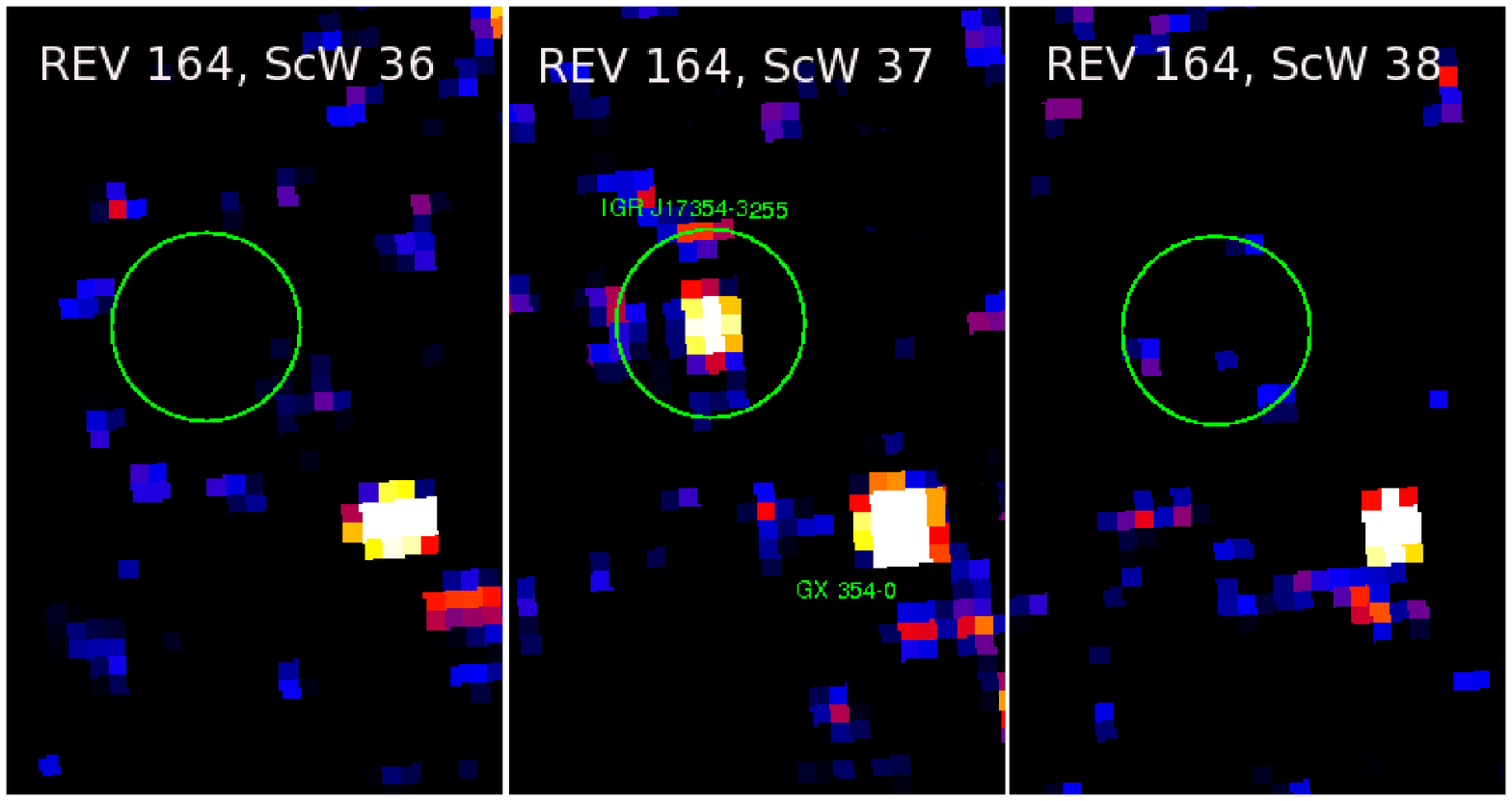, height=4.5cm,width=8.5cm}
\caption{IBIS/ISGRI ScW image sequence (18--60 keV) of flare N. 2 in Table 1 from  IGR J17354$-$3255 (circled). 
The source was detected in the middle ScW with a significance of $\sim$ 5.5$\sigma$. A persistent source (the LMXB GX 354-0) 
is also visible in the field of view.} 
\end{figure}

\subsubsection{Periodicity analysis}
In order to search for any evidence of periodicity, we investigated the IBIS long-term 
light curve of IGR J17354$-$3255 using the Lomb-Scargle periodogram method by means of the fast implementation 
of Press \& Rybicki (1989)  and Scargle (1982).
A clear signal is seen at 8.4474 days as shown in Fig. 3. 
The 99.999\% confidence level was defined at a power of 20.214 by 
using a randomisation test with 200,00 trials, as outlined in Hill et al. (2005), 
hence the detected periodic signal is the only significant peak within 
the power spectrum. Using a variant of the randomisation test as outlined in Drave et al. (2010), 
the error on  this peak was calculated as 0.0017 days. We interpret the periodicity of 8.4474$\pm$0.0017 days
as the orbital period  of the binary system, providing a strong confirmation of the result
recently reported by D'Ai et al. (2011) with \emph{Swift}/BAT data.

Fig. 4 shows the phase folded light curve where it is evident a smooth orbital modulation of the flux; it peaks during the periastron passage and becomes consistent with zero around apastron. We note that the feature at phase $\sim$ 0.5 is due to the binning and is not taken as being of physical origin. The red crosses indicate the outbursts detected by IBIS (listed in Table 1) within these orbital ephemeris. As clearly evident their occurrence is consistent with the region of orbital phase around periastron. The shape of the orbital profile shown in Fig. 4 is rather smooth and appears to be dominated by lower level X-ray emission rather than by the outbursts. To test this assumption we employed the recurrence analysis technique which searches for periastron detections (Bird et al. 2009) by summing the X-ray emission during each periastron passage within a set window of 3 days (periastron $\pm$ 1.5 days). The source could then be detected even though a significant detection is not achieved in the individual ScWs. The same process was performed for each apastron passage and the distributions compared. Fig. 5 shows the recurrence analysis results, there is a clear excess in detections above 3$\sigma$ during the periastron passages, corresponding to detectable emission on about 26\% of periastron passages covered by the data set. This value is taken as a lower limit as there may still be emission that is occurring during other periastron passages that is below the sensitivity of IBIS/ISGRI. On the contrary no detections above 3$\sigma$ are recorded during apastron passages.

Our recurrence analysis suggests that the 16 individual outbursts detected by IBIS/ISGRI cannot explain the smooth shape seen in the phase folded light curve. Assuming a source distance of 8.5 kpc (see section 1) these outbursts all have X-ray luminosities in excess of 10$^{36}$ erg s$^{-1}$ and represent the most luminous outburst events. Hence we would not expect these events to define the orbital emission profile over the extent of these long baseline observations. Instead we attribute the shape to lower level emission that is below the instrumental sensitivity of IBIS/ISGRI in an individual 
ScW (i.e. $\sim$ 10 mCrab). However when the whole data set, covering about 300 orbital cycles of 8.4 days, is folded this emission sums to a significant detection and reveals the smooth profile shown in Fig. 4. This emission could be either smoothly varying following a similar profile during each orbit or the super position of many low intensity flares at fluxes of $\sim$ 10$^{33}$--10$^{35}$ erg s$^{-1}$. Such flaring activity was demonstrated occurring in the SFXT IGR J18483$-$0311 by Romano et al. (2009) who estimated the probability of observing the source in a non-flaring state (i.e. $<$ 10$^{33}$ erg s$^{-1}$)  as only $\sim$ 25\% through observations covering an entire orbital period with Swift/XRT. Due to the sensitivity 
limits of IBIS/ISGRI and the different observing strategy 
however it is not possible to define which of these processes is occurring in this system and draw any conclusions as to the stellar wind configurations that could be responsible for this effect.

\begin{figure}
\epsfig{file = 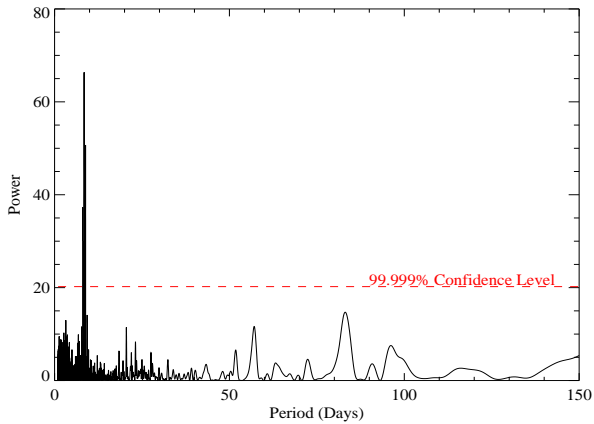, height=6cm, width=8.5cm}
\caption{Lomb-Scargle periodogram of IGR J17354$-$3255 IBIS/ISGRI light curve showing a clear signal at 8.4474 days} 
\end{figure}
\begin{figure}
\epsfig{file = 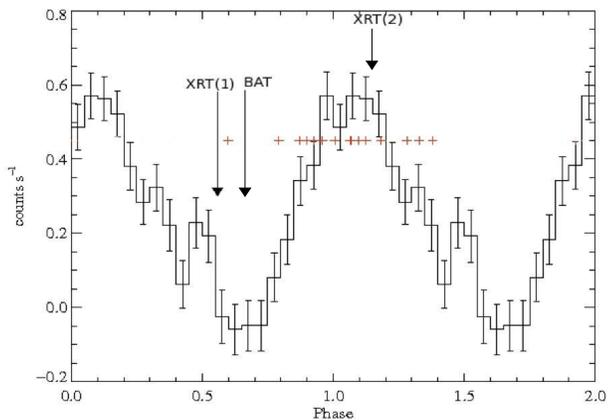, height=6cm, width=8.5cm}
\caption{Phase folded lightcurve of IGR J17354$-$3255, zero phase ephemeris MJD 52698.205. We note that the feature at 
phase $\sim$ 0.5 is due to the binning and is not taken as being of physical origin. The red crosses mark all the outbursts detected by IBIS
while the outburst detected by  \emph{Swift}/BAT  is marked by means of an arrow. Moreover, the observations during which the source was not detected
(1) and was detected (2) by \emph{Swift}/XRT are also marked by means of an arrow.} 
\end{figure}
\begin{figure}
\epsfig{file = 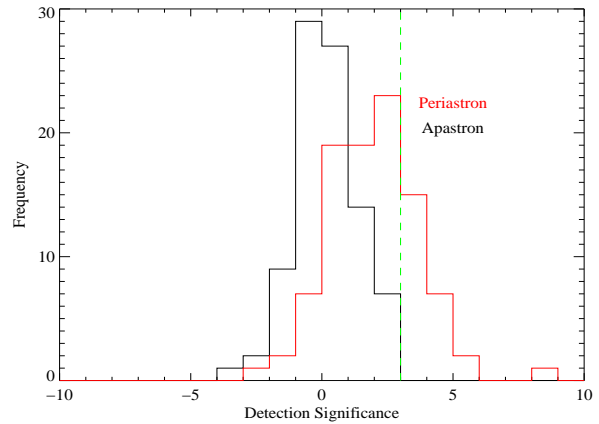, height=6cm, width=8.5cm}
\caption{Recurrence analysis for periastron and apastron detections of IGR J17354$-$3255 within 
set window of 3 days} 
\end{figure}

\subsection{\emph{Swift}}
The 15--50 keV \emph{Swift}/BAT light curve on daily timescale was inspected searching 
for outbursts. We considered only those detected with a significance 
greater than 5$\sigma$. Only one significant outburst was clearly found (see Fig. 6), peaking at MJD 55145 (10 November 2009)
with a flux of (2.8$\pm$0.54)$\times$10$^{-9}$ erg cm$^{-2}$ s$^{-1}$ or $\sim$ 220 mCrab (15--50 keV). 
The total outburst duration was $\sim$ 6 days, for a  significance detection of $\sim$ 6.5$\sigma$.
\emph{Swift}/BAT observations covered April 2009 through  May 2011 and they are not overlaped with those performed 
by INTEGRAL/IBIS which on the contrary covered February 2003 through October  2008. 
The apparent incongruence between the number of flares seen by IBIS (16)  and BAT (1) in almost the same energy band 
could be reasonably explained by the significantly different dataset as well as instrumental capabilities. 
In fact, BAT observations had an effective  exposure time  
of $\sim$ 6.4 Ms which is significantly smaller than that of INTEGRALI/IBIS ($\sim$ 10 Ms). 
In addition, Table 1 clearly shows that about 70\%  of the outbursts 
detected by IBIS/ISGRI have typical duration and flux of few hours and  20--40 mCrab, respectively. Longer flares 
(i.e. few days duration) are rarer. Clearly IBIS/ISGRI is particularly suited to the detection of short events
thanks to its high instantaneous sensitivity for short observations on $\sim$ 2000 seconds ScW lengths (i.e. $\sim$ 10 mCrab) 
which match very well the duration of the flares. On the contrary, BAT is more suited to long-term monitoring thanks to its
much more continuous coverage  but could eventually miss the detection
of shorter flares (i.e. few hours duration ) which, if close to the IBIS sensitivity, are very likely 
not detectable by BAT due to its poorer instantaneous sensitivity. In fact,  the minimum detectable 
flux by BAT depends on both the exposure and the position of the source in the sky. 
For the median BAT pointing duration of $\sim$ 1000 seconds and a source near the center of the BAT FOV, a 6$\sigma$ 
detection corresponds to  a minimum detectable flux of approximately 60 mCrab. 
In the daily averages it corresponds to  approximately 15 mCrab (Krimm et al. 2006).

Regarding \emph{Swift}/XRT, the region of the sky including IGR J17354$-$3255 was observed on 17 April 2009
($\sim$ 5.3 ks exposure). Fig. 4 clearly shows that the observation  took 
place during the periastron passage.  A  single bright X-ray counterpart was clearly detected inside the IBIS 
error circle (see inset in Fig. 7). The 0.2--10 keV light curve  (200 seconds bin time) of this X-ray counterpart is 
shown in Fig. 7. It is evident that at the beginning the source count rate was very low, after which   
several fast X-ray flares occurred and the strongest one reached a peak-flux 
of (8.7$\pm$1.1)$\times$10$^{-11}$ erg cm$^{-2}$ s$^{-1}$. Throughout the entire \emph{Swift}/XRT observation, lasting $\sim$ 8 hours, 
the source was strongly variable displaying a dynamic range of $\sim$ 50.

\begin{figure}
\epsfig{file = 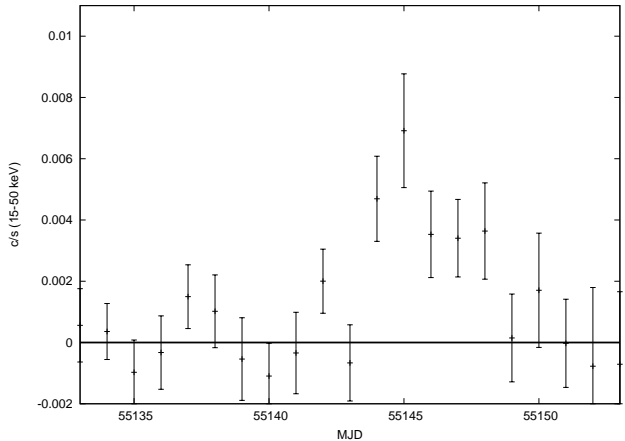, height=8.5cm,width=6cm, angle=-90}
\caption{\emph{Swift}/BAT light curve on daily timescale of the single observed outburst from IGR J17354$-$3255 (15--50 keV).} 
\end{figure}

\section{X-RAY SPECTRAL ANALYSIS}
\subsection{Flaring activity}
\subsubsection{INTEGRAL}
To date no spectral information above 20 keV  on X-ray flares from  IGR J17354$-$3255 has been available.
The number of flares reported in Table 1 allows a  spectral study for the first time. 
We were able to extract a meaningful IBIS/ISGRI spectrum only from the flares having 
a significance detection $>$ 5$\sigma$.
Due to the limited statistics we performed a 
fit by using only a  simple power law model whose spectral parameters are  listed 
in Table 1 (photon index, $\chi^{2}_{\nu}$, d.o.f.). 
The photon indices are not well constrained because of the limited statistics, however their best fit values fall 
in the  range $\Gamma$=2--3 and this
could suggest a rather constant soft spectral shape from flare to flare.  
Bearing this in mind and to improve the statistics, the IBIS/ISGRI spectra from all flares were 
summed up together and fit with a power law model ($\chi^{2}_{\nu}$=1.4, 15 d.o.f.), providing a  much better constrained photon index of 
$\Gamma$=2.4$\pm$0.4 (see Fig. 8, top spectrum). 

During just two flares (N.  8 and 9 in Table 1) the source was inside the FOV of the X-ray monitor JEM-X,  although for very short 
time ($\sim$ 1 ks and 0.5 ks, respectively) because of its  much smaller FOV with respect to that of IBIS/ISGRI.
The source was not detected,  providing a loose 3$\sigma$ upper limit of $\sim$ 5$\times$10$^{-11}$ erg cm$^{-2}$ s$^{-1}$ 
in both  3--10 and 3--20 keV energy bands.

\subsubsection{\emph{Swift/XRT}}
IGR J17354$-$3255 was detected by \emph{Swift}/XRT data on 17 April 2009
during an observation lasting $\sim$ 5.3 ks. The X-ray  spectrum 
is best fit by an absorbed power law model ($\chi^{2}_{\nu}$=1.28, 16 d.o.f) with $\Gamma$=1.7$^{+1.1}_{-0.6}$ 
and N$_H$=8$^{+4.4}_{-2.2}$ $\times$10$^{22}$ cm$^{-2}$. The expected Galactic 
absorption along the line of sight is 1.18$\times$10$^{22}$ cm$^{-2}$ (Kalberla et al. 2005).
The unabsorbed average flux is (1.6$\pm$0.14)$\times$10$^{-11}$ erg cm$^{-2}$ s$^{-1}$ (0.2--10 keV). 
In addition, an absorbed black body model  gave a reasonable description of the data as well 
($\chi^{2}_{\nu}$=1.4, 16 d.o.f) with  kT=1.4$^{+0.4}_{-0.3}$ keV 
and N$_H$=5$^{+2.2}_{-1.6}$ $\times$10$^{22}$ cm$^{-2}$. Such spectral parameters are in fair agreement 
with those previously reported by using the same \emph{Swift}/XRT dataset (Vercellone et al. 2009, D'Ai et al. 2011).

\begin{figure}
\epsfig{file = 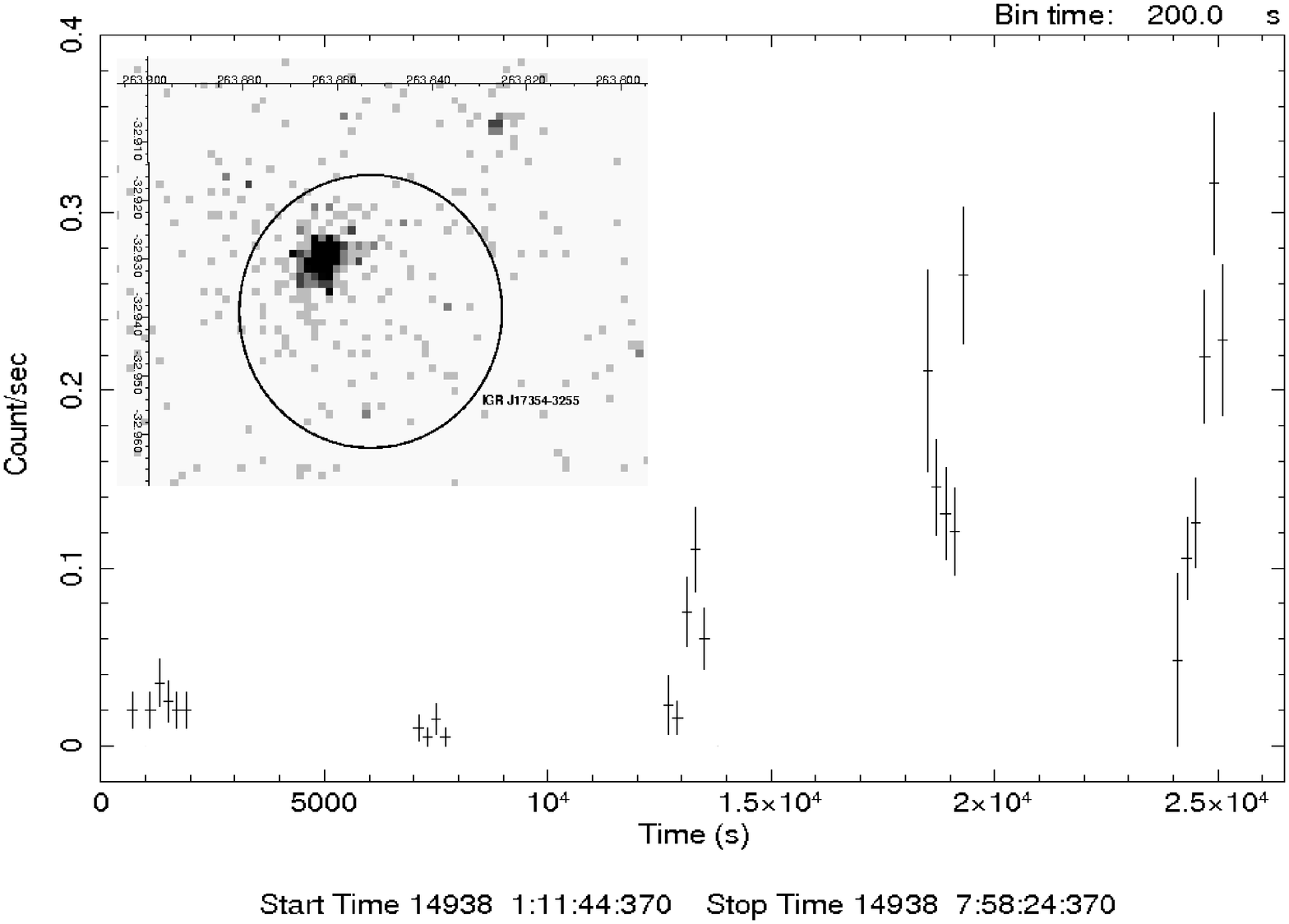,height=9cm,width=6cm,angle = -90}
\caption{\emph{Swift}/XRT light curve of IGR J17354$-$3255 (0.2--10 keV, 200 seconds bin time) during the observation on 17 April 2009. The
inset shows the IBIS/ISGRI error cicle of IGR J17354$-$3255 superimposed on the \emph{Swift}/XRT image (0.2--10 keV).} 
\end{figure}

\subsection{Out-of-outburst X-ray emission}
\subsubsection{INTEGRAL}
With the aim of investigating the hard X-ray state of IGR J17354$-$3255 outside its  bright 
flaring behaviour, we selected  all the available ScWs
during which the source was within the fully coded FOV of IBIS/ISGRI with a significance value
less than 4$\sigma$ (18--60 keV), i.e. not undergoing a bright outburst.  
The relative extracted average spectrum (see Fig. 8, bottom spectrum)
is fit by a power law  with $\Gamma$=2.4$\pm$0.4 ($\chi^{2}_{\nu}$=0.6, 5 d.o.f.). 
Fig. 8 clearly shows the source spectral constancy in shape, but not in flux, from flaring state (top) to out-of-outburst emission (bottom).
The average 18--60 keV flux is (1.4$\pm$0.1)$\times$10$^{-11}$ erg cm$^{-2}$ s$^{-1}$ or 1.1$\pm$0.1 mCrab, 
such measurement represents the lowest detectable hard X-ray state of the source to date. When assuming the lowest and highest 
source flux in outburst, as measured by INTEGRAL/IBIS and \emph{Swift}/BAT, 
we can infer a dynamic range in the interval  20--200. 

Next, we also used all the available ScWs during which the source was within the fully coded FOV 
of JEM-X1 and JEM-X2. As in the previous case, we intentionally excluded those pointings  during which it was in outburst. 
A mosaic significance map was generated in the energy band  3--20 keV, for a total on-source
exposure of $\sim$ 650 ks (JEM-X1) and $\sim$ 330 ks (JEM-X2). IGR J17354$-$3255 was not detected by JEM-X2 
providing a 3$\sigma$  upper limit of $\sim$ 2$\times$10$^{-12}$ erg cm$^{-2}$ s$^{-1}$ (3--20 keV). Conversely it was weakly detected by JEM-X1 
at $\sim$ 6$\sigma$ level, likely thanks to the longer time interval analyzed, with an average flux of 
$\sim$  1$\times$10$^{-12}$ erg cm$^{-2}$ s$^{-1}$ (3--20 keV).  Unfortunately, the insufficient 
JEM-X statistics did not allow us to extract a meaningful spectrum and light curve of this out-of-outburst state.

\subsubsection{\emph{Swift}/XRT}
From analysis of archival \emph{Swift}/XRT data,  IGR J17354$-$3255 was not detected 
during an observation on 11 March 2008 lasting $\sim$ 4.4 ks.  Fig. 4 clearly shows that it took 
place very close to the  apastron passage. We infer a 3$\sigma$ upper limit of
$\sim$ 2.8$\times$10$^{-13}$ erg cm$^{-2}$ s$^{-1}$ (0.2--10 keV)  by assuming the same spectral model as in 
the \emph{Swift}/XRT detection on 2009. When considering the highest source flux as measured by \emph{Swift}/XRT in the 
same energy band, we can infer a dynamic range $>$311.

\begin{figure}
\epsfig{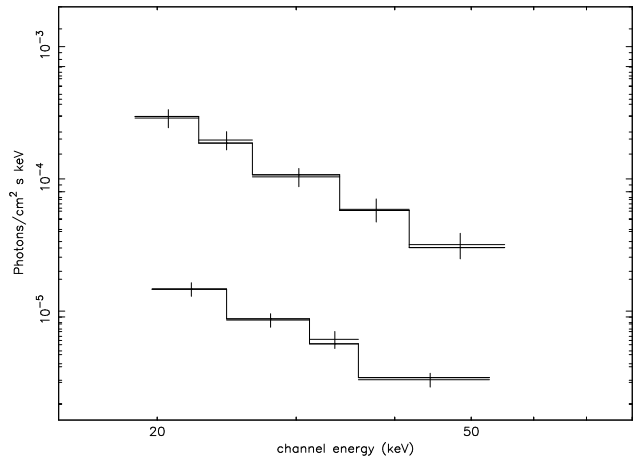}
\caption{Unfolded average 18--60 keV spectrum of IGR J17354$-$3255  during the outburst (top) and out-of-outburst emission (bottom). 
Both spectra, fit with a simple power law model, have been rebinned for display purposes.} 
\end{figure}

\section{Association with AGL J1734$-$3310?}
AGL J1734$-$3310 is an unidentified transient MeV source discovered by AGILE on 14 April 2009 
during a flare lasting only  one day (Bulgarelli et al. 2009). It was detected at $\sim$ 4.5$\sigma$ level  
only in the energy band 100--400 MeV with a flux of $\sim$ 3.5$\times$10$^{-6}$ photons  cm$^{-1}$ s$^{-1}$ and 95\% confidence error circle radius of
0$^\circ$.65 (Bulgarelli et al. 2009). 
We note that three hard X-ray sources listed in the 4th IBIS catalog (IGR J17354$-$3255, GX 354$-$0, 4U 1730$-$335)
are spatially associated with AGL J1734$-$3310 in view of its 
large positional uncertainty.

After the discovery of AGL J1734$-$3310, extensive searches for further flaring  gamma-ray emission 
have been carried out by the AGILE team.  As result, several additional MeV flares have been discovered in the AGILE data archive 
related to the period 2007--2009,  all of them have a similar duration (about one day) and gamma-ray 
fluxes in  the range (1.3--2.3)$\times$10$^{-6}$ photons  cm$^{-1}$ s$^{-1}$  
(Bulgarelli et al. in preparation,  private communication). This clearly shows that AGL J1734$-$3310 is a recurrent 
transient MeV source. The sum of all flaring episodes detected by AGILE allowed the determination of a source position 
centered   at Galactic coordinates l= 355$^\circ$.805 and b= -0$^\circ$.26, with a significance detection of 6.6$\sigma$ 
and with statistical and systematic error radius of 0.46 degrees (95\% confidence).
We note that now IGR J17354$-$3255 is the only hard 
X-ray source unambigously located within such refined AGILE position. This is clearly evident in Fig. 9 which 
shows the 18--60 keV IBIS/ISGRI significance mosaic map ($\sim$7.6 Ms exposure) of the sky 
region surrounding IGR J17354$-$3255 with superimposed the refined positional uncertainty of AGL J1734$-$3310. 
Moreover, we point out that IGR J17354$-$3255 is also the only soft X-ray source (3--10 keV) detected 
by  JEM--X1 ($\sim$ 650 ks exposure) to be located inside the AGILE error circle. The spatial 
association between IGR J17354$-$3255 and AGL J1734$-$3310 is further strengthened by our reported findings
which show for the first time that IGR J17354$-$3255 occasionally 
displays hard X-ray flares  on short timescale, i.e. a temporal behaviour similar to that  observed by AGILE 
from AGL J1734$-$3310.

\begin{figure}
\epsfig{file = 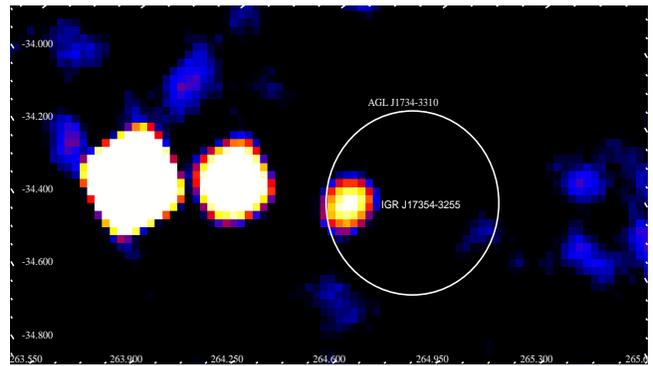,height=4.8cm,width=8.5cm}
\caption{IBIS/ISGRI mosaic significance map (18-60 keV, $\sim$7.6 Ms exposure time) of the sky region including IGR J17354$-$3255. The error circle 
(radius 0.46 degrees) represents the MeV source AGL J1734$-$3310. The other two bright sources detected in the field are the LMXBs 
GX 354--0 and 4U 1730--335.}
\end{figure}

\section{DISCUSSIONS}
In this work  we report for the first time detailed  temporal  X-ray results on 
IGR J17354$-$3255. The long-term INTEGRAL monitoring allows us to show 
that it is a weak persistent hard X-ray source spending  a major fraction of the time during 
out-of-outburst X-ray state with average 18--60 keV flux of $\sim$1.4$\times$10$^{-11}$  erg cm$^{-2}$ s$^{-1}$. 
Occasionally, the source undergoes short 
flaring activity whose duration
rangies from  a few hours (0.5--5 h) to a few days (1--3 d) while the peak-fluxes  are in the interval 
(3--28)$\times$10$^{-10}$ erg cm$^{-2}$ s$^{-1}$ (18--60 keV).
Hence,  the highest inferred dynamical range is $\sim$ 200. 

In the soft X-ray band (0.2--10 keV), 
the dynamic range of IGR J17354$-$3255
from non-detection 
to highest level of activity 
is $>$311.  When active, the source is strongly variable (factor of $\sim$ 50) on timescales of a few hours.

Our above findings, both at soft and hard X-rays, strongly resemble those of Supergiant Fast X-ray Transients (SFXTs), 
a new subclass of supergiant HMXBs mainly discovered by INTEGRAL (Sguera et al. 2005, 2006) and characterized by short 
and bright X-ray flares on top of longer and fainter periods of X-ray emission level. In this respect, IGR J17354$-$3255 could well 
belong  to the SFXT class.  However we point out that  whereas classical SFXTs have a remarkable dynamical range of $\sim$ 10$^3$--10$^4$, 
we found for IGR J17354$-$3255 significantly lower values, in the range 21--200 (18--60 keV) and $>$311 (0.2--10 keV). 
Although IGR J17354$-$3255  cannot be considered a classical SFXT, still its dynamical range is significantly 
greater than that of classical  persistent supergiant HMXBs known to display variability of factors lower than $\sim$ 20 (Walter \& Zurita 2007). 
Therefore we suggest that IGR J17354$-$3255  is an intermediate SFXT system,  much like several similar 
cases reported in the literature (Clark et al. 2010, Walter \& Zurita 2007, Sguera et al. 2007). 

In addition, our IBIS timing analysis of the source  identified an orbital period of 8.4474$\pm$0.0017 days,  
strongly confirming  the previous detection with \emph{Swift}/BAT (D'Ai et al. 2011). 
The occurrence of the outbursts detected by IBIS is consistent with the region of orbital phase around periastron. 
The calculated lower limit on the recurrence rate of the source (i.e. detected outbursts when 
predicted by the $\sim$ 8.4 days period)  is $\sim$ 26\%, this is the second highest value of any known SFXT after that of 
SAX J1818.6$-$1703 ($\sim$50\%,  Bird et al. 2009). We suggest that IGR J17354$-$3255 is a SFXT 
system similar to SAX J1818.6$-$1703 (Bird et al. 2009),  but with a shorter orbit  and a 
lower eccentricity of $\sim$ 0.1--0.2 to account for the reduced  recurrence rate.
The shape of the orbital profile  is very smooth, conversely  that of other known SFXTs is sharper. 
Our recurrence analysis suggests that the sixteen individual outbursts detected by IBIS/ISGRI, which all have X-ray 
luminosities in excess of 10$^{36}$ erg s$^{-1}$ and represent the most luminous outburst events, 
cannot explain such smooth shape. Instead we attribute it to a lower level x-ray emission that is below the instrumental sensitivity of IBIS/ISGRI in an individual ScW. This emission could be either smoothly varying following a similar profile during 
each orbit or the super position of many low intensity flares at X-ray luminosities  of $\sim$ 10$^{33}$ -- 10$^{34}$ erg s$^{-1}$.
Unfortunately, due to the sensitivity limits of IBIS/ISGRI in such short period, it is not possible to define which of these processes 
is occurring in this system and draw any conclusions.

We note that the $\sim$ 8.4 days orbital period of IGR J17354$-$3255 is significantly 
longer than those typical of Low Mass X-ray Binaries ($<$0.5 d) and it is significantly 
shorter than those of Be HMXB systems ($>$ 30 days). On the contrary,  it is 
more typical of SGXBs (2--12 d) providing further strong support to the hypothesis 
that the companion donor is a supergiant star.  In this respect, only one bright 2MASS 
counterpart is located inside the Chandra error circle of IGR J17354$-$3255
while there are no catalogued USNO optical objects. Spectroscopy of this infrared counterpart is essential to fully characterize it.
The high inferred optical extinction and the relatively high X-ray column density suggest that IGR J17354$-$3255
is located at a relatively large distance, i.e. near the Galactic Center region (d$\sim$8.5 kpc).
If we assume such a distance, then the source displays typical outbursts X-ray luminosities of 10$^{36}$--10$^{37}$ erg  s$^{-1}$ 
(18--60 keV) while its lowest X-ray luminosity state is $<$2.4$\times$10$^{33}$ erg  s$^{-1}$ (0.2--10 keV). 
Such values are  very similar to those of known firm SFXTs. 

Our deep INTEGRAL significance mosaic maps  
both at soft (3--10 keV) and  hard X-rays (18--60 keV), combined with a refined  AGILE positional uncertainty, 
revealed that IGR J17354$-$3255  is the only X-ray source unambigously 
located inside the error circle of the unidentified transient MeV source AGL J1734$-$3310. This spatial association is enforced 
by our temporal findings which unveiled for the first time the recurrent X-ray 
flaring behaviour of IGR J17354$-$3255 on short timescale (from few hours to few days), i.e. a characteristic very similar to that observed 
with AGILE from AGL J1734$-$3310.  We took into account the possibility of  a chance coincidence and 
to this aim we calculated the probability of finding a supergiant HMXB, such as IGR J17354$-$3255, inside the AGILE error circle by chance. 
Given the number of supergiant HMXBs detected by IBIS within the Galactic plane (Bird et al. 2010), defined here as restricted to a latitude range of $\pm$5 degrees, we estimated a probability of $\sim$ 1\%,  i.e. $\sim$ 0.5 chance coincidences are expected. 

It is noteworthy that the possible association between IGR J17354$-$3255 and AGL J1734$-$3310 might not be a unique and 
rare case. So far three HMXBs (LS 5039, LSI +61 303, PSR B1259$-$63) have been unambiguously 
detected both at MeV and TeV energies as persistent and  variable sources, where the variability is modulated by the 
orbital period (Paredes 2008, Hill et al. 2010). In additon, the two HXMBs Cygnus X-3 and Cygnus X-1 
have been detected at MeV energies as transient sources displaying  fast flares with a short duration of 
a few hours/days (Sabatini et al. 2010, Tavani et al. 2009, Abdo et al. 2009).
Different theoretical models have been proposed to explain the emission mechanism 
at such high energies in HMXB systems (Bosch-Ramon et al. 2006, Paredes et al. 2006, Romero et al. 2003, 2005, 
Maraschi \& Treves 1981,  Tavani \& Arons 1997). 
Similarly, SFXTs have all the ingredients to possibly be MeV/TeV emitters since they host a compact object 
(black hole or neutron star)  and a massive supergiant star which could be the source of seed photons (for the inverse Compton emission) 
and target nuclei (for hadronic interactions). However, due to their transitory nature, MeV/TeV emission from  SFXTs must 
be in the form of fast flares which are  not easy to detect. Despite this, a few SFXTs have been proposed in the literature 
as best candidate counterparts of unidentified transient MeV sources located on the Galactic Plane
(Sguera et al. 2009a, 2009b). In this respect, we propose the candidate intermediate SFXT IGR J17354$-$3255  as best  candidate counterpart of 
AGL J1734$-$3310.   Although we are aware that the reported  evidences are  
so far circumstantial, the eventual confirmation of a common nature 
would have very major implications: the SFXTs could represent a new class of MeV/GeV emitting Galactic transients.
To this aim, further and deeper multiwavelength studies of IGR J17354$-$3255 and AGL J1734$-$3310
are strongly needed especially in  radio and at gamma-rays. In particular, a periodicity analysis of the MeV emission 
from AGL J1734$-$3310 is essential to support a secure identification 
with IGR J17354$-$3255. These  kind of temporal studies can be performed by the current gamma-ray instrumentations, 
such as Fermi/LAT or AGILE, as amply proved by the LAT detections of periodic MeV emission from several HMXB systems (e.g. Hill et al. 2010).
On-going regular and frequent scannings of the Galactic Bulge and Plane with INTEGRAL, 
together with observations from Fermi/LAT and AGILE, could likely shed new light on the proposed association.

\section*{Acknowledgments}
We thank the referee for useful comments which improved the quality of the paper.
VS is very grateful to Andrea Bulgarelli and the AGILE team for sharing the results on AGL J1734$-$3310 before publication. 
The italian authors acknowledge the ASI financial support via grant ASI-INAF I/033/10/0 and I/009/10/0.
S. P. Drave acknowledges support from the Science and Technology Facilities Council STFC. 
This research has made use of the IGR sources page maintained by J. Rodriguez \& A. Bodaghee (http://irfu.cea.fr/Sap/IGR-Sources/) and  
of data obtained from the High Energy Astrophysics Science Archive Research Center
(HEASARC), provided by NASA’s Goddard Space Flight Center. It has also made use of public access \emph{Swift}-BAT light curves
provided by the \emph{Swift}-BAT team (http://swift.gsfc.nasa.gov/docs/swift/results/transients).

\end{document}